# Topological dragging of solitons

Yaroslav V. Kartashov,* Victor A. Vysloukh,** and Lluis Torner

*ICFO-Institut de Ciencies Fotoniques, and Universitat Politecnica de Catalunya, 08034, Barcelona, Spain*

We put forward properties of solitons supported by optical lattices featuring topological dislocations, and show that solitons experience attractive and repulsive forces around the dislocations. Suitable arrangements of dislocations are even found to form soliton traps, and the properties of such solitons are shown to crucially depend on the trap topology. The uncovered phenomenon opens a new concept for soliton control and manipulation, e.g., in disk-shaped Bose-Einstein condensates.

*PACS numbers: 42.65.Tg, 42.65.Jx, 42.65.Wi*

Optical lattices are unique objects that have found various applications in many branches of the fundamental science. Among the most representative examples are Bose-Einstein condensates trapped in standing light waves [1], the possibility of arrangement, guidance or deflection of microscopic particles with reconfigurable lattices [2], fabrication of photonic crystals by holographic lithography [3], etc. In optics such lattices can be used to create refractive index modulations that substantially affect properties of the light beams propagating in nonlinear media [4], opening broad prospects for creation of guiding structures whose properties can be fine-tuned by the lattice-inducing waves. The periodic optical lattices formed with standing or several interfering plane waves can support diverse types of the matter wave and optical solitons, including one- and two-dimensional, multi-humped, and vortex solitons [1,4,5]. Such lattices, in addition, offer the possibility to manage diffraction broadening of beams and an enriched set of phenomena [6], previously encountered only in technologically fabricated waveguide arrays (for a recent review see Ref. [7]).

The properties of solitons supported by the lattice are dictated to a large extent by the shape of the lattice-creating wave. Besides sets of plane waves that are capable to produce periodic lattices, one can form lattices with nondiffracting beams, such as Bessel beams. Bessel lattices exhibit radial symmetry that offers new opportunities for soliton existence and control in comparison with periodic ones [8]. Yet, all optical lattices investigated so far feature a topologically regular structure. However, there exists a whole completely distinct class of lattices possessing *higher topological complexity*, e.g., lattices containing dislocations. In contrast to regular lattices, the guiding properties of such lattices can change dramatically in the vicinity of dislocations, a possibility that leads to important new physical phenomena. In this respect, notice the recently demonstrated creation of a wide variety of optical potentials for atom manipulation with spatial light modulators [9-11]. Soliton propagation in topologically nontrivial networks of evanescently coupled waveguides and star-shaped optical networks created in Bose-Einstein condensates have been also analyzed very recently [12].

In this Letter we put forward the concept that solitons can experience dragging forces when released into lattices containing topological dislocations. Such phenomenon induces a soliton dragging along predetermined paths that are defined by the type of the dislocation. We reveal that several dislocations form soliton traps supporting stationary solitons whose properties and stability depend crucially on trap topology. To be specific, we consider lattices produced by the interference between plane waves and waves with nested vortices that can be imprinted in disk-shaped Bose-Einstein condensates, but the idea can be extended to other topological landscapes and physical settings wherever stable lattices with topological dislocations can be formed.

The generic equation describing the evolution of matter (optical) wave packets in the presence of the mean field (Kerr-type) cubic nonlinearity and weak optical potential (transverse refractive index modulation) is the nonlinear Schrödinger equation:

$$i\frac{\partial q}{\partial \xi} = -\frac{1}{2}\left(\frac{\partial^2 q}{\partial \eta^2} + \frac{\partial^2 q}{\partial \zeta^2}\right) - q|q|^2 - pR(\eta,\zeta)q. \qquad (1)$$

In the matter wave context $q$ stands for the dimensionless mean field wave function of two-dimensional condensate; $\eta, \zeta$ are measured in units of the fluid healing length; $\xi$ is the normalized time. Parameter $p$ is proportional to lattice depth expressed in units of the recoil energy [1,11]. Here we assume that optical lattice features the intensity of the interference pattern of plane wave and wave carrying one or several nested vortices. In the simplest case of single-vortex wave $R(\eta,\zeta) = |\exp(i\alpha\eta) + \exp(im\phi + i\phi_0)|^2$, where $\alpha$ is the plane wave propagation angle, $\phi$ is the azimuthal angle, $m$ is the topological charge of the vortex, and $\phi_0$ defines the orientation of vortex origin in transverse plane. In the case of multiple vortices the last term in expression for $R(\eta,\zeta)$ should be replaced with a product of several functions describing vortices with charges $m_k$, positions $\eta_k, \zeta_k$, and orientations $\phi_k$. Such optical lattices may be produced with spatial light modulators [9,10] that were successfully used for optical confinement of Bose-Einstein condensates [11]. Notice, that the lattice produced by the interference of plane wave and wave with several nested vortices is not stationary and distorts upon propagation. However, it may be implemented in two-dimensional, or disk-shaped Bose-Einstein condensates, which are elongated in transverse $(\eta,\zeta)$ plane and strongly confined in the direction of light propagation ($\rho$ axis), so that the lattice-creating waves does not vary inside the condensate thickness. Such strong confinement in Bose-Einstein condensates can be achieved in asymmetric magnetic harmonic traps with the ratio of oscillation frequencies $\nu_\rho / \nu_{\eta,\zeta} \sim 100$, where $\nu_\eta = \nu_\zeta$ (see Refs. [13] for the details of the experimental realization of disk-shaped condensates in magnetic and optical traps). Further we neglect the impact of harmonic trapping potential in Eq. (1) because of the considerable difference in the characteristic transverse scales of this potential and lattice $R(\eta,\zeta)$. Indeed, the parameters of the lattice produced with spatial light modulator can be made close to that for harmonic lattices imprinted with standing light waves [1,9-11]. Thus, typical transverse scale of lattice $R(\eta,\zeta)$ in Eq. (1) may range from 1 to 10 $\mu$m. An illustrative estimate for lithium condensates [14] with ground-state scattering length $a_s \approx -14.5$ Å, trapped in the potential with frequency $\nu_\rho = 150$ Hz, yields condensate thickness in the direction $\rho$ of strong confinement $\sim 3$ $\mu$m that is much smaller than characteristic length of lattice distortion.

In the optical context $q$ is the dimensionless field amplitude; the longitudinal $\xi$ and transverse $\eta, \zeta$ coordinates are scaled to the diffraction length and the input beam width, respectively; parameter $p$ is proportional to refractive index modulation depth. Complex refractive index landscapes can be induced in suitable optical crystals, e.g. by femtosecond laser pulses [15], a possibility that might allow exploration of our concept in such setting too. Optically-induced lattices might be another possibility, provided that dynamical instabilities of the lattice can be overcome (e.g., by introducing additional periodicities, as in [16], or stabilizing nonlinear effects). Equation (1) admits some conserved quantities including the soliton norm (or energy) $U = \int \int_{-\infty}^{\infty} |q|^2 \, d\eta d\zeta$.

Some representative examples of lattices produced by interference of plane wave and wave with nested unit-charge vortex are shown in Fig. 1. We set $\alpha = 4$ and vary $\phi_0$ and $p$. Such lattices feature clearly pronounced transverse channels where solitons can travel. The screw phase dislocation results in appearance of topological defects in the form of the characteristic fork, so that some lattice channels can fuse or disappear in the dislocation. Notice, that lattice periodicity is broken only locally, in a close proximity of dislocation, while far from the dislocation lattice profile resembles that of the quasi-one-dimensional lattice with modulation along $\eta$ axis that could be produced by interference of two plane waves $\exp(\pm i\alpha\eta)$. Far from the dislocation the period of the lattices shown in Fig. 1 along $\eta$ axis is given by $T = 2\pi/\alpha$. The lattice structure in the dislocation depends on the orientation of vortex origin $\phi_0$ and vortex charge. The higher the charge the stronger the lattice topological complexity and more channels can fuse or disappear in the dislocation.

The central result of this Letter is that lattice topological features produced by screw phase dislocations profoundly affect behavior of solitons launched into the lattice, so that lattice intensity gradient in the vicinity of the dislocation results in appearance of forces acting on the soliton. The sign and modulus of these forces are dictated by the type of dislocation and, in particular, by the charge of screw phase dislocation in lattice-creating wave. The interaction strength decreases with increase of the distance between soliton center and dislocation. Close to the dislocation forces acting on soliton results in its fast transverse displacement when it is initially launched parallel to $\xi$ axis. Figure 1 shows directions of displacement for different initial soliton positions. We used as an input solitons with $U = 4.5$ supported by undistorted quasi-one-dimensional lattice with period $T = 2\pi/\alpha$ and placed them into channels of distorted lattice close to dislocation. In this case soliton profile remains almost unchanged during evolution while its center experiences progressive shift in the transverse plane. For both vortex orientations shown in Fig. 1 solitons start to move initially in negative direction of $\zeta$-axis. Solitons that are launched at $\zeta > 0$ can be bounced back at bends of lattice channels when they approach dislocation (this is indicated by double arrows). Notice, that direction of forces acting on solitons can be reversed by the inversion of vortex charge (this is accompanied by the corresponding lattice modification). Since far from the dislocation solitons move freely along undistorted channels of the lattice, it is possible to introduce asymptotic escape velocity $\alpha_{\text{esc}}$ for solitons released at position $(0, -\delta\zeta)$. The escape velocity $\alpha_{\text{esc}}$ decreases with increase of initial separation between soliton and dislocation (Fig. 1).

The dependence of sign and magnitude of forces acting on solitons on the lattice topology comprises the possibility to use combinations of several dislocations (produced by spatially separated oppositely charged vortices) for creation of double lattice defects, or *soliton traps*, which feature *equilibrium* regions that are capable to support stationary solitons. Several examples of such traps are shown in Fig. 2. Positive traps are formed when origins of oppositely charged vortices are oriented in opposite directions (positive and negative directions of $\zeta$ axis) and several lattice rows fuse to form region where $R(\eta, \zeta)$ is locally increased. The negative traps are produced when the vortex origins are oriented in the same direction and feature the break-up of several lattice channels, i.e. regions where $R(\eta, \zeta)$ is locally decreased. A qualitative map of forces acting on solitons in the vicinity of the traps can be obtained with the aid of effective particle approach [6] that yields the equations

$$\begin{aligned} \frac{d^2}{d\xi^2}\langle\eta\rangle &= \frac{p}{U}\int_{-\infty}^{\infty}\int_{-\infty}^{\infty}|q|^2\frac{\partial R}{\partial\eta}\,d\eta\,d\zeta, \\ \frac{d^2}{d\xi^2}\langle\zeta\rangle &= \frac{p}{U}\int_{-\infty}^{\infty}\int_{-\infty}^{\infty}|q|^2\frac{\partial R}{\partial\zeta}\,d\eta\,d\zeta, \end{aligned} \qquad (2)$$

for soliton center coordinates $\langle\eta\rangle = U^{-1}\int\int_{-\infty}^{\infty}\eta|q|^2\,d\eta d\zeta$, $\langle\zeta\rangle = U^{-1}\int\int_{-\infty}^{\infty}\zeta|q|^2\,d\eta d\zeta$. We constructed the map of forces $F_\eta \sim d^2\langle\eta\rangle/d\xi^2$ and $F_\zeta \sim d^2\langle\zeta\rangle/d\xi^2$ for a simplest Gaussian ansatz $|q| = q_0\exp[-2(\eta^2 + \zeta^2)]$ for the soliton profile (other ansatz produce qualitatively similar results).

Figure 3 shows the map of the $\zeta$-component of the force defining the direction of soliton motion along lattice channels for the case of the traps created by wave with two nested vortices with charges $m = \pm 1$ and separation $\delta\zeta$. Both positive and negative traps repel solitons located in the lattice channels outside the trap (some of the possible soliton positions are indicated by the circles in Fig. 3). The force $F_\zeta$ changes its sign at $\zeta = 0$ that indicates a possibility of existence of stationary solitons in this region. We found them numerically from Eq. (1) in the form $q(\eta,\zeta,\xi) = w(\eta,\zeta)\exp(ib\xi)$, where $b$ is the propagation constant (or chemical potential in the matter wave case) and $w(\eta,\zeta)$ is the real function. The typical profiles of solitons localized at the positive and negative traps are shown in Fig. 4. Depending on the trap length and type (positive or negative) the profile of soliton can be elongated or compressed along the $\zeta$ axis. For small length of the positive trap the energy of the corresponding solitons is a monotonically growing function of propagation constant (Fig. 5(a)) and vanishes at the cutoff $b_{\rm co}$. Close to the cutoff the soliton expands over several lattice channels. At $b \to \infty$ the transverse extent of soliton decreases and its energy approaches the critical value $U_{\rm cr} \approx 5.85$ encountered for unstable solitons in the absence of the lattice. With increase of the trap length above a certain value, the $U(b)$ dependence changes qualitatively (see Fig. 5(b)) and gradually approaches dependence $U_{\rm h}(b)$ encountered for solitons supported by periodic quasi-one-dimensional lattice. Figure 5(c) shows $U(b)$ diagram for solitons supported by negative lattice trap. In this case soliton ceases to exist in the cutoff without any topological shape transformation, while its energy decreases monotonically with $b$. Following the Vakhitov-Kolokolov stability criterion one may expect that positive lattice traps are able to support stable solitons with $dU_{\rm p}/db > 0$, while solitons supported by negative lattice traps are unstable since $dU_{\rm n}/db < 0$. Validity of such criterion was fully confirmed by the extensive numerical simulations of the evolution Eq. (1) in the presence of random perturbations to ideal soliton shapes. This result confirms the possibility to generate desired soliton trapping potentials by using the adequate topological landscapes. It also confirms the crucial importance of trap topology and its transverse extent for properties and stability of solitons that it supports, as well as for magnitude and sing of forces experienced by outermost solitons.

    In conclusion, we put forward the new concept of dragging of solitons by optical lattice featuring topological dislocations. We revealed that lattice gradients induced by dislocations produce attractive and repulsive forces on solitons released into the lattices. As a consequence, we uncovered that several dislocations can even form traps supporting localized solitons. Complex light field distributions required for creation of lattices with dislocations in disk-shaped Bose-Einstein condensates can be generated with suitable spatial light modulators.

*On leave from Physics Department, M. V. Lomonosov Moscow State University, Russia. **Departamento de Fisica y Matematicas, Universidad de las Americas, Puebla, Mexico.

# Figure captions

Figure 1. Lattices with dislocations (top) produced by the interference pattern of plane wave and wave with nested vortex for various vortex orientations. Arrows show direction of motion for solitons placed into lattice channels in the vicinity of dislocation. Soliton escape velocity versus the initial distance to dislocation (bottom). Input soliton energy $U = 4.5$. Vortex charge $m = 1$.

Figure 2. Top row shows positive (left) and negative (right) topological traps created with two vortices with charges $m = \pm 1$. Bottom row shows the same but for vortex charges $m = \pm 3$. Trap length $\delta\zeta = \pi$.

Figure 3. Map of $F_\zeta$ force for solitons launched in the vicinity of positive (left) and negative (right) topological traps. Arrows show direction of soliton motion. Trap length $\delta\zeta = \pi$.

Figure 4. Profiles of solitons supported by various topological traps. (a) Positive trap, vortex charges $m = \pm 1$, $b = 2.8$. (b) Negative trap, $m = \pm 1$, $b = 4.1$. (c) Positive trap, $m = \pm 3$, $b = 3.5$. In all cases trap length $\delta\zeta = \pi$.

Figure 5. (a) Energy vs. propagation constant for solitons supported by positive topological trap with length $\delta\zeta = \pi$ and by quasi-one-dimensional lattice. (b) Modification of $U(b)$ dependence with increase of positive trap length. (c) Energy vs. propagation constant for solitons supported by negative trap with length $\delta\zeta = \pi$.

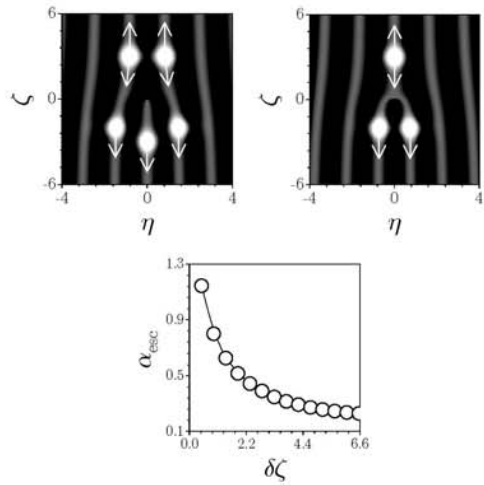

Figure 1. Lattices with dislocations (top) produced by the interference pattern of plane wave and wave with nested vortex for various vortex orientations. Arrows show direction of motion for solitons placed into lattice channels in the vicinity of dislocation. Soliton escape velocity versus the initial distance to dislocation (bottom). Input soliton energy $U = 4.5$. Vortex charge $m = 1$.

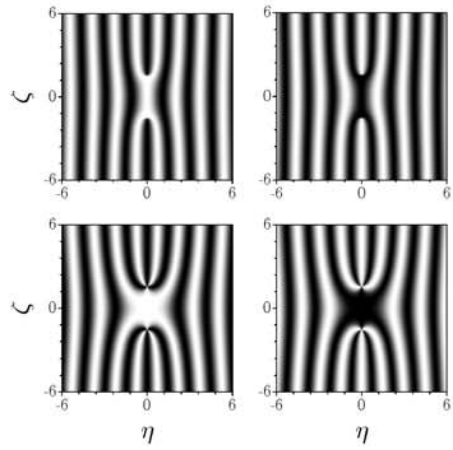

Figure 2. Top row shows positive (left) and negative (right) topological traps created with two vortices with charges $m = \pm 1$. Bottom row shows the same but for vortex charges $m = \pm 3$. Trap length $\delta\zeta = \pi$.

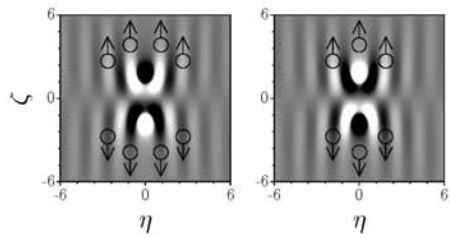

Figure 3. Map of $F_\zeta$ force for solitons launched in the vicinity of positive (left) and negative (right) topological traps. Arrows show direction of soliton motion. Trap length $\delta\zeta = \pi$.

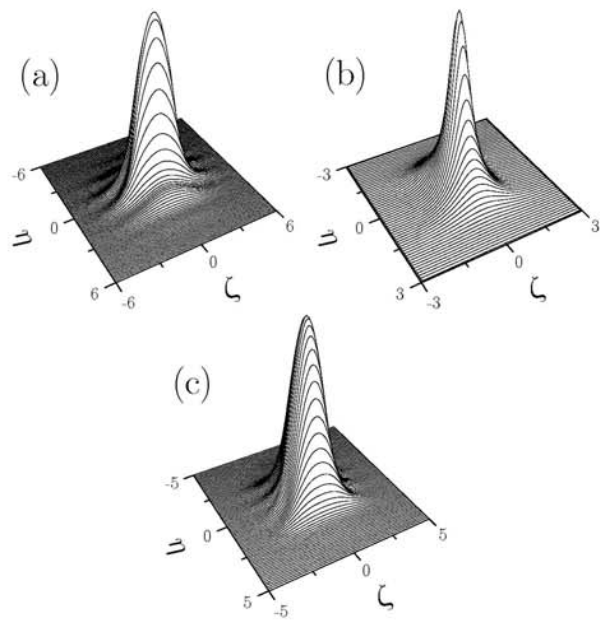

Figure 4. Profiles of solitons supported by various topological traps. (a) Positive trap, vortex charges $m = \pm 1$, $b = 2.8$. (b) Negative trap, $m = \pm 1$, $b = 4.1$. (c) Positive trap, $m = \pm 3$, $b = 3.5$. In all cases trap length $\delta\zeta = \pi$.

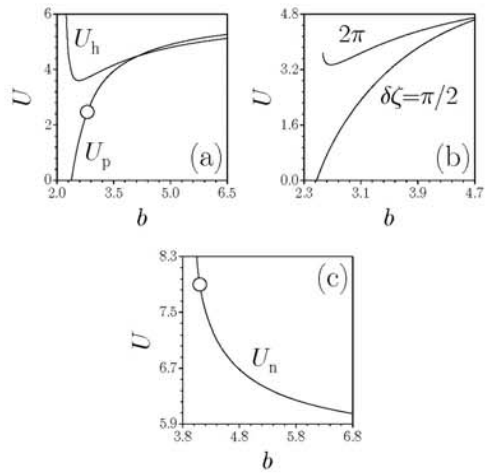

Figure 5.  (a) Energy vs. propagation constant for solitons supported by positive topological trap with length $\delta\zeta = \pi$ and by quasi-one-dimensional lattice. (b) Modification of $U(b)$ dependence with increase of positive trap length. (c) Energy vs. propagation constant for solitons supported by negative trap with length $\delta\zeta = \pi$.